%% file: main.tex
\documentclass[10pt,conference,review,anonymous]{IEEEtran}
\IEEEoverridecommandlockouts

\usepackage{cite}
\usepackage{amsmath,amssymb,amsfonts}
\usepackage{graphicx}
\usepackage{textcomp}
\usepackage{xcolor} 
\usepackage{algpseudocode}
\usepackage{algorithm}
\usepackage{xspace}
\usepackage{listings}
\usepackage{amssymb}
\usepackage{color}
\usepackage{multirow}
\usepackage{dblfloatfix}
\usepackage{lscape} 
\usepackage{xparse}
\usepackage{placeins}
\usepackage{caption} 
\usepackage{subcaption}
\usepackage{enumitem}
\usepackage{array, makecell}
\usepackage{wrapfig}
\usepackage{colortbl}
\usepackage{tabularx} 
\usepackage{booktabs}
\usepackage{courier}
\usepackage{ragged2e}
\usepackage{amsthm}
\usepackage{hyperref}
\usepackage{comment}
\usepackage{pdflscape}
\usepackage{afterpage}
\usepackage{fontawesome}
\usepackage{pifont}
\usepackage{lscape}
\usepackage{enumitem}

\setlist[itemize]{noitemsep, topsep=4pt}
\definecolor{lightgray}{HTML}{fafafa}
\definecolor{darkgray}{rgb}{.55,.55,.55}
\definecolor{darkblue}{HTML}{0066cc}
\definecolor{brickred}{HTML}{b04f4f}
\definecolor{purple}{rgb}{0.65, 0.12, 0.82}
\definecolor{diffadd}{HTML}{00b359} 
\definecolor{diffrmbg}{HTML}{ffebe9}
\definecolor{diffaddbg}{HTML}{e6ffeb}
\definecolor{diffremove}{HTML}{de4f54}
\definecolor{carrotorange}{rgb}{0.8, 0.33, 0.0}
\definecolor{highlight}{HTML}{fefbc2}
\definecolor{bluegray}{HTML}{3182bd}
\definecolor{lightred}{HTML}{3182bd}

\lstdefinelanguage{JavaScript}{
  keywords={typeof, new, true, false, catch, function, return, null, catch, switch, var, const, let, extends, if, in, while, do, else, case, break, async, await, of},
  keywordstyle=\color{darkblue}\bfseries,
  ndkeywords={class, export, boolean, throw, implements, import, this, setTimeout},
  ndkeywordstyle=\color{brickred}\bfseries,
  identifierstyle=\color{black},
  sensitive=false,
  comment=[l]{//},
  morecomment=[f][\color{diffadd}\bfseries]{+\ },
  morecomment=[s]{/*}{*/},
  morecomment=[f][\color{diffremove}\bfseries]{- },
  commentstyle=\color{violet}\ttfamily,
  stringstyle=\color{carrotorange}\ttfamily,
  morestring=[b]',
  morestring=[b]"
}

\lstdefinelanguage{Python}{
  keywords={typeof, new, true, false, catch, function, return, null, catch, switch, var, const, let, extends, if, in, while, do, else, case, break, async, await, of, from, import, class, def},
  keywordstyle=\color{darkblue}\bfseries,
  ndkeywords={class, export, boolean, throw, implements, import, this, setTimeout, self, __init__},
  ndkeywordstyle=\color{brickred}\bfseries,
  identifierstyle=\color{black},
  sensitive=false,
  comment=[l]{//},
  morecomment=[f][\color{diffadd}\bfseries]{+\ },
  morecomment=[s]{/*}{*/},
  morecomment=[f][\color{diffremove}\bfseries]{- },
  commentstyle=\color{violet}\ttfamily,
  stringstyle=\color{carrotorange}\ttfamily,
  morestring=[b]',
  morestring=[b]"
}

\lstdefinelanguage{Dockerfile}{
  keywords={FROM, RUN, CMD, LABEL, MAINTAINER, EXPOSE, ENV, ADD, COPY, ENTRYPOINT, VOLUME, USER, WORKDIR, ARG, ONBUILD, STOPSIGNAL, HEALTHCHECK, SHELL},
  keywordstyle=\color{darkblue}\bfseries,
  ndkeywords={--from, --chown, --mount},
  ndkeywordstyle=\color{brickred}\bfseries,
  identifierstyle=\color{black},
  sensitive=true,
  comment=[l]{\#},
  morecomment=[f][\color{diffadd}\bfseries]{+\ },
  morecomment=[f][\color{diffremove}\bfseries]{- },
  commentstyle=\color{darkgray}\ttfamily,
  stringstyle=\color{violet}\ttfamily,
  morestring=[b]',
  morestring=[b]"
}

\theoremstyle{definition}

\newcommand{\header}[1]{\par\smallskip\noindent\textbf{#1.}}

\AtBeginDocument{%
  }

\newboolean{showcomments}
\setboolean{showcomments}{true}
\ifthenelse{\boolean{showcomments}}
{
	\definecolor{myyellow}{RGB}{255, 228, 26}
	\definecolor{myblue}{RGB}{50, 50, 220}
	\newcommand{\nb}[2]{
		{\sf
			\fcolorbox{myyellow}{yellow}{\scriptsize\textbf{#1}}%
			$\blacktriangleright$%
			{\color{myblue}\fontsize{7pt}{8pt}\selectfont\textbf{#2}}%
		}%
	}
}
{
	\newcommand{\nb}[2]{}
}

\newcommand{\toolname}{\textsc{FlakiDock}\xspace}
\newcommand{\datasetname}{\textsc{Flake4Dock}\xspace}

\newcommand{\parfum}{\textsc{Parfum}\xspace}
\newcommand{\shipwright}{\textsc{Shipwright}\xspace}
\newcommand{\code}[1]{{\small\ttfamily\texttt{#1}}}

\newcommand{\gptthree}{\textsc{GPT-3.5}\xspace}
\newcommand{\gpt}{\textsc{GPT-4}\xspace}

\newcommand{\shipwrightrepos}{20,526\xspace}
\newcommand{\studiedrepos}{18,055\xspace}
\newcommand{\candidaterepos}{8,132\xspace}
\newcommand{\noncandidaterepos}{9,923\xspace}
\newcommand{\flakydockerfiles}{798\xspace} 
\newcommand{\flakydockerfilespercentage}{9.81\%\xspace}
\newcommand{\flakybuilds}{974\xspace} 
\newcommand{\nonflakydockerfiles}{7,325\xspace} 
\newcommand{\repaireddockerfiles}{100\xspace}

\begin{document}

\title{Dockerfile Flakiness: Characterization and Repair}









\makeatletter
\newcommand{\linebreakand}{%
  \end{@IEEEauthorhalign}
  \hfill\mbox{}\par
  \mbox{}\hfill\begin{@IEEEauthorhalign}
}

\makeatother
\author{
     \IEEEauthorblockN{Taha Shabani}
     \IEEEauthorblockA{
         \textit{University of British Columbia}\\
         Vancouver, Canada \\
         taha.shabani@ece.ubc.ca
     }
     \and
     \IEEEauthorblockN{Noor Nashid}
     \IEEEauthorblockA{
         \textit{University of British Columbia}\\
         Vancouver, Canada \\
         nashid@ece.ubc.ca
     }
     \linebreakand
     \IEEEauthorblockN{Parsa Alian}
     \IEEEauthorblockA{
         \textit{University of British Columbia}\\
         Vancouver, Canada \\
         palian@ece.ubc.ca
     }
     \and
     \IEEEauthorblockN{Ali Mesbah}
     \IEEEauthorblockA{
         \textit{University of British Columbia}\\
         Vancouver, Canada \\
         amesbah@ece.ubc.ca
     }
}

\maketitle
\thispagestyle{plain}
\pagestyle{plain}

\input{abstract}


\maketitle

\input{introduction}
\input{motivation}

\input{characteristics}
\input{proposed-approach}
\input{evaluation}
\input{discussion}

\input{threats}

\input{related-work}
\input{conclusion}

\bibliographystyle{IEEEtran}
\interlinepenalty=10000
\bibliography{references}

\end{document}

%% file: abstract.tex
\begin{abstract}

Dockerfile flakiness—unpredictable temporal build failures caused by external dependencies and evolving environments—undermines deployment reliability and increases debugging overhead. Unlike traditional Dockerfile issues, flakiness occurs without modifications to the Dockerfile itself, complicating its resolution. In this work, we present the first comprehensive study of Dockerfile flakiness, featuring a nine-month analysis of 8,132 Dockerized projects, revealing that around 10\% exhibit flaky behavior. We propose a taxonomy categorizing common flakiness causes, including dependency errors and server connectivity issues. Existing tools fail to effectively address these challenges due to their reliance on pre-defined rules and limited generalizability. To overcome these limitations, we introduce \toolname, a novel repair framework combining static and dynamic analysis, similarity retrieval, and an iterative feedback loop powered by Large Language Models (LLMs). Our evaluation demonstrates that \toolname achieves a repair accuracy of 73.55\%, significantly surpassing state-of-the-art tools and baselines.
\end{abstract}

%% file: introduction.tex
\section{Introduction}
\label{sec:introduction}
Docker~\cite{docker} is a set of platform-as-a-service (PaaS) products that use OS-level virtualization to automate building, deploying, and delivering applications in containers. 
This approach simplifies deployment across diverse systems and plays a critical role in the CI/CD pipeline, highlighting the importance of ensuring Dockerfile reliability during builds to maintain software dependability.


Existing academic \cite{giovanni:not-all-dockerfile-smells-are-same:msr24, henkel:binnacle:icse2020, zhou:drive:arXiv2022, durieuxparfum:parfum:2024, rosa:fixing:arXiv2022, shipwright:icse2021, dockercleaner:icsm2023, eng:msr2021} and industrial efforts mainly focus on identifying Dockerfile smell or bug patterns. 
Static analysis studies~\cite{henkel:binnacle:icse2020, hadolint, dockerfilelint, dockercleaner:icsm2023} offer a linter to detect smells inside Dockerfiles. 
While existing techniques effectively identify various Dockerfile smells, they often rely on pre-defined rules that are frequently outdated or poorly maintained \cite{dockerfilelint}. \parfum \cite{durieuxparfum:parfum:2024}  enriches the Dockerfile AST with structural information derived from command lines, enabling the automatic detection and repair of smells. 
%
%
\shipwright~\cite{shipwright:icse2021} employs a human-in-the-loop approach to repair broken Dockerfiles by clustering failure patterns using a modified BERT model \cite{reimers2019sentence} and HDBSCAN \cite{HDBSCAN}, then formalizing solutions into a repair pattern database with regular expressions and fix functions.

%

%

Unlike test flakiness~\cite{lue:flakiness:fse2014}, Dockerfile builds can exhibit \emph{temporal failures} that emerge over time due to dynamic changes in external dependencies, such as updates to base images, third-party libraries, or environmental settings---occurring without any modifications to the Dockerfile itself.
These temporal failures can manifest as either deterministic or non-deterministic behaviours, depending on the nature of the underlying issues. We refer to this phenomenon as \emph{Dockerfile flakiness}---the unpredictable temporal behaviour of the Dockerfile build process,  where builds may fail over time even though the content of the Dockerfile itself remains unchanged. Deterministic failures consistently cause builds to fail after a specific point, often due to changes such as the deprecation of a base image or the removal of a required dependency. In contrast, non-deterministic failures occur sporadically, resulting in random build successes and failures. Such issues are typically caused by transient factors, including server connectivity problems or temporary inconsistencies in package manager caches.

Existing studies \cite{giovanni:not-all-dockerfile-smells-are-same:msr24, henkel:binnacle:icse2020, zhou:drive:arXiv2022, durieuxparfum:parfum:2024, rosa:fixing:arXiv2022, shipwright:icse2021, dockercleaner:icsm2023, hadolint, dockerfilelint} assume the reliability of the Dockerfiles and their build output over time, so that a build failure indicates the changes applied directly to the source Dockerfile.
The literature largely overlooks the issue of flakiness in Dockerfile builds, 
which can disrupt the deployment process, particularly in CI/CD pipelines, by hindering automatic builds. This leads to delays and consumes valuable developer time and effort in diagnosing and resolving the underlying causes of the flakiness.

In this work, we first examine the consistency and reliability of Dockerfile builds over time. 
In a longitudinal study spanning a nine-month timeframe, we built and analyzed Dockerfiles of \candidaterepos open-source projects and observed that \flakydockerfiles (\flakydockerfilespercentage) exhibit flakiness behavior. Based on the observed instances of flakiness, we developed a taxonomy for characterizing Dockerfile flakiness. Our findings indicate that the most common types of flakiness are related to dependencies, web server connectivity, and security/authentication issues. 

Second, we observe that existing techniques, such as \parfum and \shipwright, exhibit significant limitations in repairing flaky Dockerfiles. \parfum primarily addresses code smells, making it less effective in tackling flakiness issues. \shipwright, on the other hand, faces challenges including: (1) heavy reliance on human intervention, which hinders scalability and introduces variability due to dependence on the supervisor’s expertise, (2) limited generalizability of repair patterns to diverse or unseen errors, (3) reliance on pattern-matching and regular expressions, which can introduce inconsistencies, and (4) the need for ongoing maintenance of the repair database to accommodate emerging failure types.

To overcome these shortcomings, we present \toolname, a novel approach that leverages feedback-directed retrieval-augmented generation (RAG) with large language models (LLMs) to automatically repair Dockerfile flakiness. By integrating static and dynamic analysis, similarity retrieval, and an iterative LLM-based feedback loop, \toolname effectively resolves complex flakiness issues.


In this work, we make the following contributions:

\begin{itemize}

\item The first study of Dockerfile flakiness characterization. We present the first taxonomy of Dockerfile flakiness by building Dockefiles and analyzing build outputs of \candidaterepos Dockerized projects.

\item A dataset called \datasetname including \flakydockerfiles and \nonflakydockerfiles flaky and non-flaky Dockerfiles. Flaky Dockerfiles are accompanied by categorization and build errors. We also provide repair information for \repaireddockerfiles flaky Dockerfiles, specifically designed to evaluate flakiness detection and repair tasks.

\item An iterative LLM-based technique called \toolname that leverages both static and dynamic information from Dockerfiles to repair flakiness.

\item An empirical evaluation of \toolname, assessing its effectiveness with Dockerfile repair tools and baselines such as \gpt. 
\end{itemize}

Our results show that \toolname achieves a 73.55\% repair accuracy in resolving Dockerfile flakiness, significantly outperforming \parfum (0.58\%), \shipwright (5.52\%), and GPT-4-based prompting with build output (37.79\%). 



%% file: motivation.tex
\section{Motivation}
\label{sec:motivation}


Listing \ref{lst:mot1} shows a flaky Dockerfile, the corresponding build failure, and a subsequent repair based on the build error. According to the build output (lines 95 to 114), using a virtual environment is required to install \code{pip} for the specified \code{alpine} base image in Dockerfile (line 1) due to the adaptation of \code{Python Enhancement Proposal (PEP) 668}~\cite{pep668}, which addresses Externally Managed Environments. This specification prevents package managers such as \code{pip} from modifying packages in the interpreter’s default environment, ensuring compatibility and reducing the risk of breaking the underlying operating system managed by external package managers. As such, this Dockerfile fails to build, whereas previous builds prior to this adaptation were all successful. The solution here would be to create and activate a virtual environment, as shown in the Dockerfile context (lines 11 and 12).



While adhering to best practices~\cite{henkel:binnacle:icse2020, hadolint, dockerfilelint, durieuxparfum:parfum:2024, dockercleaner:icsm2023} is essential to mitigate errors and vulnerabilities in Dockerfiles and in some cases to prevent potential failures from happening, we argue that this alone is insufficient to address flakiness. For example, existing tools such as HadoLint~\cite{hadolint} recommend pinning the exact version of the base image (e.g.,  rule: DL3006) or dependencies (e.g., rules: DL3007, DL3008, DL3013, DL3016, DL3018) to prevent errors caused by their internal changes. This practice can be applied to Listing \ref{lst:mot1} by using old or outdated \code{alpine} images as a solution. However, it does not provide a viable solution for the problem; applying such rules without considering the static and dynamic nature of Dockerfiles can introduce other types of flakiness, such as outdatedness and compatibility issues in the future.

\begin{lstlisting}[language=Dockerfile, captionpos=t, title={\small\textbf{Build Output}}]
...
\end{lstlisting}
\vspace{-\baselineskip}
\begin{lstlisting}[language=Dockerfile, firstnumber=87]
> [5/5] RUN pip3 install -r requirements.txt:
error: externally-managed-environment
\end{lstlisting}
\vspace{-0.9\baselineskip}
\begin{lstlisting}[language=Dockerfile, firstnumber=95, backgroundcolor=\color{highlight}]
    If the package in question is not packaged already (and
    hence installable via "apk add py3-somepackage"), please
    consider installing it inside a virtual environment, e.g.:
    ...
\end{lstlisting}
\vspace{-0.9\baselineskip}
\begin{lstlisting}[language=Dockerfile, firstnumber=114, backgroundcolor=\color{highlight}]
hint: See PEP 668 for the detailed specification.
\end{lstlisting}
\vspace{-0.9\baselineskip}
\begin{lstlisting}[language=Dockerfile, firstnumber=115]
ERROR: process "/bin/sh -c pip3 install -r requirements.txt" did not complete successfully: exit code: 1
\end{lstlisting}
\vspace{-0.9\baselineskip}
\begin{lstlisting}[language=Dockerfile, captionpos=t, backgroundcolor=\color{highlight}, title={\small\textbf{Repaired Dockerfile}}]
FROM alpine:latest
\end{lstlisting}
\vspace{-0.9\baselineskip}
\begin{lstlisting}[language=Dockerfile, firstnumber=2]
RUN apk add --update python3 py3-pip git tcpdump
RUN git clone https://github.com/649/Memcrashed-DDoS-Exploit.git Memcrashed
WORKDIR Memcrashed
...
\end{lstlisting}
\vspace{-0.9\baselineskip}
\begin{lstlisting}[language=Dockerfile, firstnumber=10, backgroundcolor=\color{highlight}]
- RUN pip3 install -r requirements.txt
+ RUN python3 -m venv venv
+ RUN . venv/bin/activate && pip install -r requirements.txt
\end{lstlisting}
\vspace{-0.9\baselineskip}
\begin{lstlisting}[language=Dockerfile, firstnumber=13, label={lst:mot1}, captionpos=b, caption={Base Image Internal Change}]
ENTRYPOINT ["python3", "Memcrashed.py"]
\end{lstlisting}

Listing \ref{lst:mot2} demonstrates another flaky Dockerfile that clings to the version pinning rule for the base image. As depicted in the Dockerfile (line 1), although the base image version is explicitly mentioned, inconsistent behavior is plausible due to using a relatively old base image. This flakiness is evident inside the build output (line 132), where the expression \code{pre\_go17} is located. The error stems from the compatibility issue of a stale \code{GOLANG} base image, i.e., older than 1.17, with existing dependencies utilized in the Dockerfile (line 8), failing the compilation and build of the project. Accordingly, a base image version upgrade is required (line 2). Furthermore, updated \code{GOLANG} images require a different approach for handling executables (lines 3 and 4) due to the adoption of new techniques.

\begin{lstlisting}[language=Dockerfile, captionpos=t, title={\small\textbf{Build Output}}]
...
\end{lstlisting}
\vspace{-\baselineskip}
\begin{lstlisting}[language=Dockerfile, firstnumber=130]
> [build-env 4/4] RUN cd /src && go build -ldflags "-linkmode external -extldflags -static" -o proxy:
\end{lstlisting}
\vspace{-0.9\baselineskip}
\begin{lstlisting}[language=Dockerfile, firstnumber=132, backgroundcolor=\color{highlight}]
/go/src/golang.org/x/net/context/pre_go17.go:47:2: background redeclared in this block
...
\end{lstlisting}
\vspace{-0.9\baselineskip}
\begin{lstlisting}[language=Dockerfile, firstnumber=153]
ERROR: process "/bin/sh -c cd /src && go build -ldflags \"-linkmode external -extldflags -static\" -o proxy" did not complete successfully: exit code: 2
\end{lstlisting}
\vspace{-0.9\baselineskip}
\begin{lstlisting}[language=Dockerfile, backgroundcolor=\color{highlight}, captionpos=t, 
title={\small\textbf{Repaired Dockerfile}}]
- FROM golang:1.9.1 AS build-env
+ FROM golang:1.22 AS build-env
+ WORKDIR /src
+ RUN go mod init my_module
\end{lstlisting}   
\vspace{-0.9\baselineskip}
\begin{lstlisting}[language=Dockerfile, firstnumber=5]
RUN go get -d -v github.com/armon/go-socks5
...
\end{lstlisting}   
\vspace{-0.9\baselineskip}
\begin{lstlisting}[language=Dockerfile, backgroundcolor=\color{highlight}, firstnumber=8]
RUN cd /src && go build -ldflags "-linkmode external -extldflags -static" -o proxy
\end{lstlisting}   
\vspace{-0.9\baselineskip}
\begin{lstlisting}[language=Dockerfile, firstnumber=9]
# final stage
FROM scratch
WORKDIR /app
...
\end{lstlisting}   
\vspace{-0.9\baselineskip}
\begin{lstlisting}[language=Dockerfile, firstnumber=15, label={lst:mot2}, captionpos=b, caption={Compatibility Issues With Stale Base Image}]
CMD ["./proxy"]
\end{lstlisting}


Dockerfiles depend on various elements such as operating systems, packages, environments, commands, and project source code, making them susceptible to diverse forms of flakiness. Listings \ref{lst:mot1} and \ref{lst:mot2} demonstrate that understanding and resolving such flaky behavior requires analyzing both the static context (Dockerfile) and dynamic context (build output) along with its temporal changes. 

%% file: characteristics.tex
\section{Characterization of Dockerfile Flakiness}
\label{sec:characteristics}


In this section we present our longitudinal investigation of Dockerfile builds, which addresses the following research questions:

\begin{itemize}
    \item \textbf{RQ1}: How prevalent is flakiness in Dockerfiles?

    \item \textbf{RQ2}: What are the main categories of Dockerfile flakiness?
\end{itemize}

The entire analysis, including project checkouts and Dockerfile builds, was conducted on an infrastructure comprising four Intel Xeon 2.50GHz machines, each with 62 GB of RAM.


\subsection{Data Collection}

For our study, we started with the \shipwright dataset \cite{shipwright:icse2021}, which contains \shipwrightrepos Docker projects with ten or more stars from GitHub repositories. The dataset includes Docker projects created up to June 2020 and focuses exclusively on projects with a single Dockerfile located in the root directory. 
We cloned the most up-to-date version of all the repositories from the \shipwright dataset. However, some repositories were no longer publicly accessible, had been removed from GitHub, or no longer contained the root Dockerfiles. Consequently, our initial pool of projects comprised \studiedrepos repositories. 

In our study, we analyze Dockerfiles along with their build outputs, which we generate by building the Dockerfiles within our infrastructure. Given that some Docker projects can be time-consuming to build, we set a 30-minute build timeout for each repository. This resulted in 93\% of builds completing without a timeout. To ensure the reliability and efficiency of our large-scale Docker build system, we use the \texttt{docker build} command with the \texttt{--no-cache} option to eliminate failures stemming from cached Docker data. Additionally, we develop a systematic Docker cleaning technique to prevent environmental and internal errors, ensuring a fresh Docker environment. To enhance time efficiency and prevent internal errors, we clean the Docker system after every four consecutive builds—a frequency determined through trial and error. The cleaning process involves removing all the cache, images, and any other peripheral leftover data during the builds alongside uninstalling and reinstalling Docker with a steady version to prevent working with a corrupted Docker system. This pipeline ensures the freshness of the Docker system throughout the builds. Using this approach, we stored 32 rounds of builds for the initial Docker projects, capturing Dockerfile build outputs in our infrastructure machines from April 2023 to December 2023.

\subsection{Flakiness Extraction}
\label{subsec:flakiness-extraction}
The detection of flakiness within test suites has been extensively explored in the prior work~\cite{ipflakies:python-order-dependent-flaky-tests:icse22, shi:ifixflakies-order-dependent-flaky-tests:fse2019}. These efforts revolve around static and dynamic test code analysis, pattern recognition, rerunning tests multiple times, and checking code changes through times in repositories. Considering the complexity and variety of Dockerfile commands, external dependencies, and configurations, directly checking the Dockerfile context for flakiness detection would require an extensive endeavor. Furthermore, due to the dynamic nature of Dockerfiles, characterizing flakiness utilizing only the static analysis approaches would fail to address all aspects.

\header{Filtering Phase} To start our flakiness study, we initiated an \emph{in-context} Docker build for all the \studiedrepos Dockerized GitHub projects in our dataset that completed the build within 30-minutes. This resulted in \candidaterepos (45.04\%) successful and \noncandidaterepos (54.96\%) failed Docker builds. We consider only the successful ones as valid candidates for our flakiness analysis due to their initial stability. 

Over a period of nine months, we rebuilt the \candidaterepos candidate projects weekly totaling 32 times. The reason for conducting builds over an extended timeframe was two-fold. First, building Dockerfiles is a time-consuming operation. The average time to build a single Dockerfile in our dataset is around eight minutes. Despite distributing our candidates across four different machines with similar operating systems and configurations and employing multiprocessing on each machine to enhance build speed, some degree of delay was unavoidable. Second, we expected that such an extended timespan could allow us to investigate temporal fluctuations in project stability. 


\header{Pre-Processsing Phase}
Dockerfile build outputs can range from a few lines to thousands, detailing the progress of each execution step. Diagnosing and extracting critical failure points from these outputs is essential but challenging, especially when error logs are lengthy or the failure point is distant from the error's manifestation. For instance, the \code{RUN pip install -r requirements.txt} command can produce hundreds of lines of output, often cluttered with warnings or unrelated messages, obscuring the root cause. To address this, we implemented a rule-based pre-processing approach that divides build outputs into stages, each corresponding to a specific Dockerfile command. Each stage includes lines containing the stage number, execution time, and log information. By analyzing each stage, we capture lines with error-related expressions like \code{ParseError} or \code{ERR!}, along with adjacent lines sharing the same execution time, to provide additional context for diagnosing failures.

\subsection{Flakiness Categorization}
A taxonomy of Dockerfile flakiness is critical for systematically addressing inconsistencies and complexities, enabling developers to identify and resolve related challenges. We employ similarity analysis to remove duplicates or identical errors within Dockerfile build outputs. This is followed by a detailed examination of pre-processed outputs, leveraging \gptthree to interpret and summarize the results. Finally, manual analysis is performed to address potential inaccuracies. 

\header{Clustering Phase} During the rebuild period, we encountered 12,964 failing build outputs while building the candidate projects. To streamline our analysis and minimize redundancy within Dockerfile build outputs, we conduct a clustering process using sentence similarity assessment within each project. To identify unique build errors in a single project, we measure the cosine similarity between the build output embeddings, extracted from the \code{all-mpnet-base-v2}~\cite{sentence-transformers} sentence transformer, 
which is trained over 1 billion sentences in different domains. Using the similarity scores, we cluster them into distinct groups. In our clustering approach, each new build output is evaluated for its similarity to existing clusters. Based on the average similarity with all current clusters, we decide whether to incorporate it into an existing cluster or create a new one. We apply this clustering method within individual projects rather than across all projects due to the intricate nature of Dockerfiles and their build outputs, which complicates the accurate clustering of errors when applied globally. To improve accuracy, we use pre-processed build outputs for similarity comparison rather than raw outputs. This approach resulted in an 87\% reduction in failing build outputs, leaving 1,684 remaining, simplifying subsequent analysis steps.


\header{Labeling Phase} We employed \gptthree to extract error descriptions encountered during builds. Given the Dockerfile context alongside the corresponding build error, the model is prompted to extract a list of contributing factors to the error and an initial label indicating the category of error. We then used these build errors, alongside the information generated by the language model, to construct a taxonomy of Dockerfile build flakiness. Our objective was to create a comprehensive hierarchical classification that captures the diverse and dynamic behavior of Dockerfiles. To achieve this, a brainstorming session among authors was conducted to design a pipeline for analyzing context and refining the labels suggested by the language model. Two authors reviewed and resolved discrepancies in the labels generated by the model, based on the Dockerfiles and build errors. Any differences in interpretation were discussed among the authors to reach a consensus. This systematic analysis resulted in the generation of category hierarchies, requiring approximately 220 person-hours of effort.

\begin{figure*}[ht] 
\centering
\includegraphics[width=\linewidth]{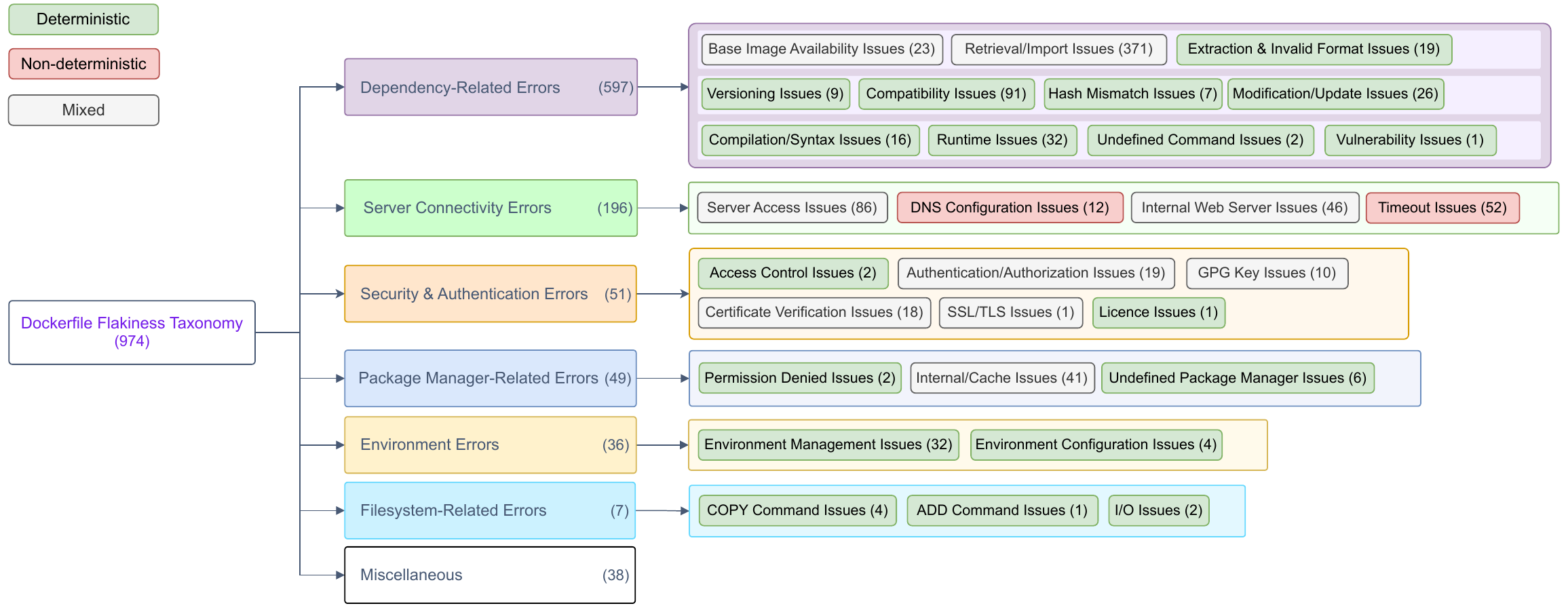}
\caption{Dockerfile Flakiness Taxonomy} 
\label{fig:taxonomy} 
\end{figure*}

\subsection{RQ1: Prevalence of Dockerfile Flakiness}
While categorizing, we selectively omitted failures stemming from our infrastructure, Docker servers, and project source code issues to hone in on genuine instances of Dockerfile flakiness. Infrastructure failures accounted for 38 of the 1,684 builds, highlighting the effectiveness of our systematic Docker cleaning method in ensuring reliability. Issues stemming from Docker servers and project-specific errors accounted for 431 and 241 failures, respectively, and were also excluded from the flaky build outputs. 

For instance, a project-specific error could involve a command such as \code{RUN build\_script.sh}, which executes a script exhibiting non-deterministic behavior due to race conditions or reliance on external services. Similarly, \code{RUN npm install} can result in project-specific errors when dependencies listed in \code{package.json} exhibit flaky or inconsistent behavior caused by misconfigurations, version mismatches, or outdated packages. In such cases, the Dockerfile itself is correct, but the failure stems from issues within the external dependency definitions.

After filtering out these non-flaky failures, we were left with \flakybuilds build outputs originating from \textbf{\flakydockerfiles (\flakydockerfilespercentage)} Dockerfiles out of \candidaterepos candidate projects, as some Dockerfiles exhibited multiple distinct flaky behaviors throughout our longitudinal analysis. This number would likely be even higher in real-world scenarios, as our candidate Dockerfiles were sourced from high-quality projects.

This percentage is significant when compared to the prevalence of flaky tests in practice. According to an empirical analysis on Flaky tests \cite{lue:flakiness:fse2014}, 4.56\% of all test failures across test executions at Google’s continuous integration (CI) system, named TAP, were reported to be due to flaky tests during a 15-month window. Another study by Microsoft \cite{lam:flakiness:issta2019} reported that 4.6\% of individual test cases monitored over a month were flaky. Our finding of \flakydockerfilespercentage flaky Dockerfiles aligns with the prevalence of flaky tests in other large-scale systems, highlighting the importance of investigating and developing tools to address Dockerfile flakiness.

\subsection{RQ2: Taxonomy of Dockerfile Flakiness}
Figure \ref{fig:taxonomy} illustrates our hierarchically structured taxonomy of Dockerfile flakiness. The left side shows the main categories of flakiness, while the right side lists the associated subcategories. The numbers within each box indicate the frequency of occurrences. 
For each subcategory, we indicate its nature—deterministic, non-deterministic, or mixed—where ``mixed" refers to cases that exhibit both deterministic and non-deterministic behaviors.
The rest of this section provides an overview of the primary categories of flakiness identified in our taxonomy. Detailed information, including examples for each category and sub-category, can be found in our replication package \cite{flakidock}.

\header{Dependency-Related Errors (DEP)} This is the most prevalent category, accounting for 61.29\% of all errors. It encompasses 11 subcategories of errors that occur during the retrieval, installation, or post-installation operations of dependencies specified in Dockerfiles. These three steps of errors are shown in the first, second, and third rows of dependency-related error subcategories in Figure \ref{fig:taxonomy}. We define a \textit{dependency} as a Base image or any external software package or library explicitly mentioned in the Dockerfile.

\textit{Availability and Retrieval Errors:} These errors happen when previously available dependencies are not found or cannot be accessed (e.g. \textsc{Retrieval/Import Issues}) 
temporarily or permanently, or decompressed due to \textsc{Extraction \& Invalid Format Issues}. As an example of \textsc{Base Image Availability Issues}, one of the Docker projects we studied, \code{Mistserver}~\cite{mistserver}
, uses \code{FROM phusion/baseimage:master}, leading to a failure because this specific version of the image is currently unavailable.

\textit{Installation Errors:} During the installation phase, flaky errors can occur due to different reasons including \textsc{Versioning Issues} or \textsc{Compatibility Issues}, where specific dependencies with altered versions or configurations may conflict or be incompatible with one another, causing the build process to fail. These errors can also involve \textsc{Hash Mismatch Issues} and \textsc{Modification/Update Issues}, where changes in dependencies lead to inconsistencies. For instance, consider a scenario where package \texttt{p1} requires package \texttt{p2} with a version greater than or equal to \texttt{v2} to be installed properly, but the existing \texttt{p2} version is older than \texttt{v2}. 

\textit{Post-Installation Errors:} After installation, flakiness can arise from the dependencies themselves, such as \textsc{Vulnerability Issues}, \textsc{Compilation/Syntax Issues}, or \textsc{Runtime Issues}. Additionally, improper installations can lead to \textsc{Undefined Command Issues} in the Dockerfile, resulting in build failures. As an example, we have observed flakiness in a Dockerfile using \texttt{Symphony}—a PHP framework—which is known to have vulnerability issues within its CodeExtension filters~\cite{codeextension}, and using those filters may cause flakiness.

\header{Server Connectivity Errors (CON)}
This category is the second most prevalent among all the categories comprising 20.12\% of the errors. These errors occur when there are issues while connecting to previously stable external servers. 
These errors are typically non-deterministic, except when servers are permanently out of service or entirely unreachable.
\textsc{Server Access Issues} arise when the Dockerfile cannot reach the target server due to invalid URLs or temporary server downtimes. \textsc{Timeout Issues} occur when connections to servers take too long to establish or complete, often caused by overloaded servers or the massive size of transferred data. \textsc{Internal Web Server Issues} refer to errors within the accessed server, typically denoted with \texttt{HTTP} status codes \texttt{500s}. \textsc{DNS Configuration Issues} arise when the Dockerfile temporarily fails to resolve the server's domain name, resulting in failed connections.

\header{Security and Authentication Errors (SEC)}
This category contains errors related to changes or deprecation of previous security protocols and authentication processes, making up 5.24\% of the total errors. \textsc{Access Control Issues} arise when the Dockerfile does not have the necessary permissions to access required internal resources caused by the base image's internal changes. \textsc{Authentication/Authorization Issues} manifest when there are problems verifying the identity of the user or service, which can result from incorrect or expired credentials or misconfigured authentication/authorization services. 

This subcategory is mixed, as such errors can be affected by transient factors like token expiration or permanent external service instability.

\textsc{SSL/TLS Issues} encompass a range of problems related to the secure transmission of data, including protocol mismatches and outdated cryptographic algorithms. \textsc{GPG Key Issues} arise when there are problems with the cryptographic keys used to verify the integrity and authenticity of downloaded items, which can prevent the successful retrieval and installation of necessary dependencies. 
The nature of these errors is mixed, as they can result from invalid keys or temporarily unavailable key servers.
Lastly, \textsc{Licence Issues} occur when the Dockerfile attempts to use software with new licensing restrictions.

\header{Package Manager-Related Errors (PMG)}
Package manager-related errors constitute 5\% of the flakiness instances and refer to the changes applied to the package manager configuration during the build process. The most common subcategory of this class is \textsc{Internal/Cache Issues} with 4.2\% of total flakiness. These errors arise from inconsistency or unreliability in the package manager's internal system during its installation or utilization, resulting in failed operations. 
This subcategory is mixed, as deterministic failures can stem from corrupted caches, while non-deterministic failures may result from transient issues with the package manager's internal operations.
As an example, the command: \code{RUN npm install --registry=r} where \code{r} is no longer a reliable registry for \code{npm} would cause an internal issue within the package manager. Another subcategory is \textsc{permission-denied issues}, which occur when the Dockerfile does not have the necessary permissions to interact with the package manager to install or update packages. This sort of error can happen due to permission changes within the base images or other infrastructures within the Dockerfile. Lastly, \textsc{Undefined package Manager Issues} encompass errors caused by improper installation of package managers, resulting in a corrupted package manager within the system.

\header{Environment Errors (ENV)}
\textsc{Environment Management Issues} and \textsc{Environment Configuration Issues} fall into this category, representing 3.7\% of the errors corresponding to interactions with virtual environments. These environment errors often arise from changes made to the base images or other underlying infrastructures specified in the Dockerfile. Such changes enforce developers to strictly adhere to the new rules to minimize vulnerabilities and enhance the system's robustness. A detailed example of this type of error is illustrated in Listing \ref{lst:mot1} where an externally managed environment is required to alleviate the risk of disrupting the OS package management system.

\header{Filesystem-Related Errors (FS)}
This category, representing the smallest portion of our study on Dockerfile flakiness, accounts for less than 1\% of the total failures. These errors include challenges in handling file system operations within Dockerfiles such as \textsc{COPY} and \textsc{ADD} command errors, and \textsc{I/O} issues generally stemming from the base image internal file system updates.

\header{Miscellaneous} During our analysis of Flaky Dockerfiles and their build outputs, we categorized 3.9\% of the instances of flakiness as Miscellaneous. This group includes builds with highly complex errors or those executed in silent mode without informative execution logs, making it challenging to pinpoint the issues and classify them.


%% file: proposed-approach.tex
\section{\toolname}
\label{sec:approach}

\begin{figure*}[ht] 
\centering
\includegraphics[width=\linewidth]{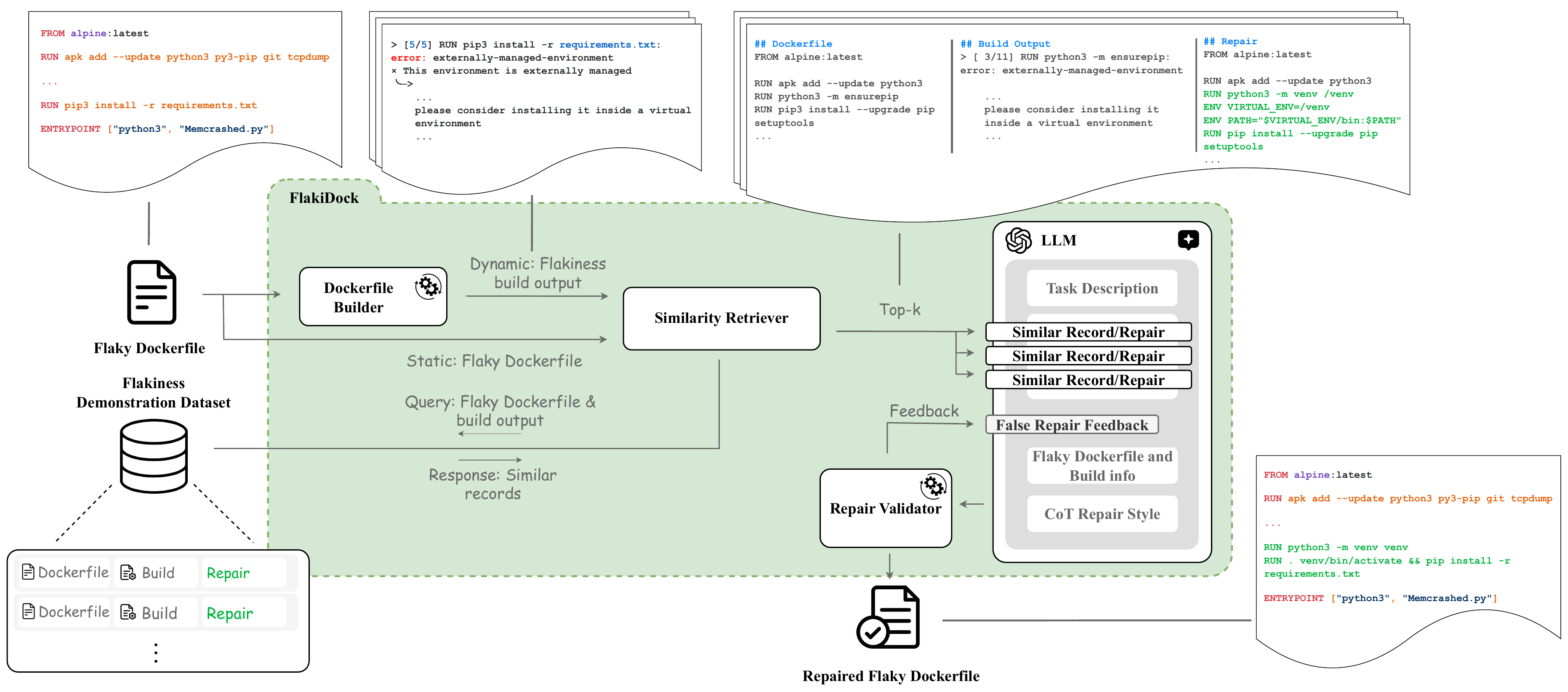}
\caption{Overview of our proposed approach: \toolname} 
\label{fig:docker-build} 
\end{figure*}

As demonstrated in Section \ref{sec:characteristics}, Dockerfile flakiness presents various complex symptoms in the build output. Leveraging the ability of LLMs to solve programming tasks across different domains~\cite{denny:conversing-with-copilot:sigcse23, ross:conversational-interaction-for-coding:iui23, xie:chatunitest:arxiv23, tian:chatgpt-programming-assistance:arxiv2023, xia:chatrepair:arxiv2023, lertbanjongam2022empirical, pearce:copilot-security:2021, nhan:copilot-empirical-evaluation:msr22, vaithilingam:copilot:chi22, dakhel2022github, saki:copilot-pair-programming:icse22}, and the effectiveness of retrieval-augmented generation (RAG) techniques~\cite{cedar:icse2023, toufique:semantic-augmentation-of-llm-prompts:icse2024, lance}, we propose \toolname, an automated approach using LLMs to repair Dockerfile flakiness. Our insight is that by providing LLMs with demonstrations containing \emph{static} (Dockerfile) and \emph{dynamic} (build outputs) information, along with repair patches from similar examples, the model can resolve flakiness in new Dockerfiles. Figure \ref{fig:docker-build} provides an overview of our approach.

\subsection{Demonstration Dataset Creation}
We first need to create a demonstration dataset for our approach. To this end, we randomly sample \repaireddockerfiles Dockerfiles from our dataset of \flakydockerfiles flaky Dockerfiles and manually provide repairs for them. To provide a robust demonstration of error categories in our repair dataset, we maintain a minimum of 30\% coverage for categories that exhibit less than 5\% distribution of Dockerfile flakiness, such as PMG, ENV, and FS categories. However, within each category, we randomly sample the Dockerfiles. If flakiness is not observed in a Dockerfile at the time of analysis, we randomly select another Dockerfile from our dataset for repair. 
We repair flaky Dockerfiles using static (Dockerfile), dynamic (build outputs), and extracted categorization information. If a repair solution is not apparent using this information, similar to \shipwright~\cite{shipwright:icse2021}, we perform a human inspection of the top five web pages from search engine results, based on querying the error keywords, to find potential solutions. 

Repairing Dockerfile flakiness is a complex and time-consuming task because it involves several intricate steps. First, one must understand the execution steps of the Dockerfile and the overall workings of the project. Then, the specific problem must be pinpointed, and finally, the suggested repair should be verified. We verified the suggested repairs using the \texttt{Docker build} command with the \code{--no-cache} option. To confirm the reliability of the proposed repairs, we test them as many times as the longest consecutive failure streak observed in our build history. Additionally, if a Dockerfile has exhibited different types of flakiness throughout the flakiness extraction phase \ref{subsec:flakiness-extraction}, we only consider the types of flakiness that can be observed during the repair step. Some instances of flakiness may be difficult to reproduce or may have been resolved due to external dependency updates. 

Table \ref{tab:repairs} shows the distribution of repairs per category. We spent approximately 115 person-hours generating \repaireddockerfiles Dockerfile repairs. Notably, the low occurrence in the server connectivity category is due to the fact that most errors in this category are temporary server issues resolved by the time of analysis.

\begin{table}[h]
    \centering
    \renewcommand{\arraystretch}{1.3} 
    \begin{tabular}{l|c|c|c|c|c|c|c}
        \toprule
        \textbf{Category} & \textbf{DEP} & \textbf{CON} & \textbf{SEC} & \textbf{PMG} & \textbf{ENV} & \textbf{FS} & \textbf{Total} \\
        \midrule
        \textbf{Repairs} & 63 & 6 & 9 & 8 & 10 & 4 & 100 \\
        \bottomrule
    \end{tabular}
    \caption{Number of Repairs by Category}
    \label{tab:repairs}
\end{table}

In our demonstration dataset, a Dockerfile is denoted with $d$, and every record is defined as tuple $(S_{d}, D_{d}, C_{d}, R_{d}, I_{d})$ containing several elements of the Flaky Dockerfile. The $S_{d}$ component contains the static information, i.e., Dockerfile context. The identifier $D_{d}$ indicates the pre-processed build output of the flaky Dockerfile, which is referred to as the dynamic information. $C_{d}$ defines the category label for the current Dockerfile based on the taxonomy illustrated in Fig \ref{fig:taxonomy}. Elements $R_{d}$ and $I_{d}$ denote the repair patch proposed for the Dockerfile and the number of iterations required for the repair to be tested to ensure its correctness, respectively. In case more than one repair is offered for a flaky Dockerfile, $R_{d}$ and $I_{d}$ form a list of values. 

\subsection{Dockerfile Building}
\label{subsection:builder-module}
The first step of \toolname is to build a given flaky Dockerfile to elicit the failing build output. The purpose of this building stage is two-fold. First, based on the build output extracted, we classify the Dockerfile as \textit{non-flaky} if no failure is detected over $n$ iterations. Conversely, if a failure occurs, we identify it as flaky behavior and proceed to address it. 
$n$ is determined based on our demonstration dataset, ensuring it is at least greater than or equal to 90\% of $I_{d}$ in our dataset, which corresponds to a minimum of two iterations. This approach is designed to encompass most flaky behaviors while optimizing time efficiency.

\subsection{Static and Dynamic Similarity Retrieval Phase}
Few-shot learning has shown remarkable efficacy when applied to LLMs~\cite{prompt-learning-survey:liu:21}. While randomly selected examples can enhance performance, recent studies \cite{cedar:icse2023, toufique:semantic-augmentation-of-llm-prompts:icse2024} indicate that choosing examples based on their similarity to the input context can lead to even more significant improvements. 

In the domain of Dockerfiles, we argue that both static and dynamic information are required to resolve flakiness. To this end, upon completing the build phase, the captured information along with the original Dockerfile is used to retrieve similar examples from the flakiness demonstration dataset. Formally, the input given to the similarity retriever is a tuple $(S_{q}, D_{q})$ where $S_{q}$ and $D_{q}$ represent the static (Dockerfile) and dynamic (build output) features respectively. Then, the retriever uses the pre-processing technique elaborated in Section \ref{subsec:flakiness-extraction} to extract error-related features from the build output. The combination of $S_{q}$ and $D_{q}$ is the input query $Q$ to the demonstration dataset. Given that our input encompasses both code segments and natural language descriptions, we utilize embedding-based search via \texttt{text-embedding-ada-002} embedding from OpenAI~\cite{ada}. This model is pretrained on extensive datasets across various domains, can process up to 8,191 tokens, and outputs a vector of 1,536 dimensions. Using this transformer model, we compare $Q$ with the records in our demonstration dataset (only $S_{e}$ and $D_{e}$) by applying cosine similarity to retrieve the top-3 $(S_{e}, D_{e})$ combinations that are closest in embedded space to the input query, along with the corresponding repair snippets $R_{e}$ for those retrieved Dockerfiles.

\subsection{Repair Generation}
This step contains the prompt design for Dockerfile flakiness repair generation. As shown in Figure \ref{fig:docker-build}, the prompt comprises a natural language task description, the flaky Dockerfile along with its build output, and a Chain-of-Thoughts (CoT) explanation to guide the model through the repair process. The prompt is then augmented with demonstration examples retrieved from the similarity retriever. Each example $e$ consists of a triple: $(S_{e}, D_{e}, R_{e})$  represents the Dockerfile, build output, and repair/repairs suggested for the flaky Dockerfile.

If previous attempts have generated incorrect repairs for the current flaky Dockerfile, a feedback message is included in the prompt as false demonstrations to help the model avoid similar mistakes. It encompasses the history of proposed repairs, each recorded as a false repair: $fr=(R_{fr}, D_{fr})$. Here, $R_{fr}$ element denotes the false repaired Dockerfile, and $D_{fr}$ shows the pre-processed build output corresponding to the false repair.

\subsection{Repair Validation}
The validation phase serves as a heuristic approach to determine whether the repair suggested by the LLM is effective. The structure of this stage is elaborated in Algorithm \ref{alg:repair-validation}. In the beginning, The repaired Dockerfile is built $n$ times, and build outputs are captured, similar to the Dockerfile builder module described in \ref{subsection:builder-module}. If all build outputs are successful, the validator confirms the repair's correctness and finalizes it as a result. Otherwise, it identifies the most common error type observed from the feedback generated thus far. The feedback comparison relies on the similarity of build outputs, assessed using sentence transformation models. 

Through a manual evaluation of our demonstration Dockerfiles, we found that after three incorrect repair attempts with the same error, LLMs tend to continue proposing flawed repairs due to hallucination or model deficiency in addressing that specific problem even with the augmented information provided. Therefore, we establish a threshold, denoted by $T$, set at a constant value of 3, to determine the stopping point for repair generation. If a specific error type appears $T$ times, we interpret it as an indication of the model's hallucination or inability to resolve the issue, resulting in the output \code{Unable to resolve!}. If no error occurs three or more times, new feedback—consisting of the false repair and its build output—is created, appended to the existing feedback list, and then incorporated into the LLM's prompt to generate a new Dockerfile flakiness repair.


\begin{algorithm}
\caption{Repair Validation}
\label{alg:repair-validation}
\scriptsize
\begin{algorithmic}
\State \textbf{Input:} 
    \State $r_{d} \gets$ Repaired Dockerfile,
    \State $f_{d} \gets$ Previous feedbacks
    \State $it \gets$ n
    \State $T \gets 3$ \Comment{Threshold}
\State \textbf{Output:} Repair / Feedback / "Unable to Resolve!"
\end{algorithmic}
\begin{algorithmic}[1]
\State $buildOutput \gets$ getBuildResults($r_{d}$, $it$)
\If {allSuccessful($buildOutput$)}
    \State \textbf{Return:} $r_{d}$ \Comment{Repair}
\Else
    \State $failures \gets countSimilarFailures(buildOutput, f_{d})$
    \If {$failures \geq T$}
        \State \textbf{Return:} \texttt{"Unable to resolve!"}
    \Else
        \State $f_{d} \gets appendNewFeedback(r_{d}, buildOutput)$
        \State \textbf{Return:} $f_{d}$ \Comment{Feedback}
    \EndIf
\EndIf
\end{algorithmic}
\end{algorithm}








%% file: evaluation.tex
\section{Evaluation}
\label{sec:evaluation}
To assess the effectiveness of \toolname we address the following research question:



\begin{itemize}    
    \item \textbf{RQ3}: How effective is \toolname and how does it compare to state-of-the-art techniques? 


    

\end{itemize}

\subsection{Implementation}
\toolname is developed in Python. For our experiments, we use the \gpt model (\texttt{gpt-4-0613}) as the LLM, with the temperature parameter set to 0 to ensure deterministic and well-defined outputs. Chroma DB, which is an open-source embedding database~\cite{chroma}, serves our need for vector similarity search. For embedding generation, we use \code{text-embedding-ada-002} from OpenAI~\cite{ada}. The temporal analysis, including project checkouts and Dockerfile builds, is performed on an infrastructure consisting of 4 Intel(R) Xeon(R) CPU 2.50GHz machines with 62 GB RAM each. For repair, we use 8 AWS machines of type t2.2xlarge, each with 8 CPUs and 32 GB RAM.

\begin{table*}[h]
    \scriptsize
    \centering
    \caption{Results for Dockerfile Flakiness Repair}
    \begin{tabular}{l | c | c | c | c | c | c | c}
        \toprule
        \textbf{Tool and Strategy} & \textbf{DEP} & \textbf{CON} & \textbf{SEC} & \textbf{PMG} & \textbf{ENV} & \textbf{FS} & \textbf{Total} \\
        \midrule
        
        \parfum & 0/280 (0\%) & 1/9 (11.11\%) & 0/16 (0\%) & 1/16 (6.25\%) & 0/22 (0\%)& 0/1 (0\%) & 2/344 (0.58\%) \\

        \shipwright & 19/280 (6.79\%) & 0/9 (0\%) & 0/16 (0\%) & 0/16 (0\%) & 0/22 (0\%)& 0/1 (0\%) & 19/344 (5.52\%) \\                
        
        \midrule
        \gpt Dockerfile & 13/280 (4.64\%) & 2/9 (22.22\%) & 1/16 (6.25\%) & 4/16 (25\%) & 1/22 (4.55\%)& 0/1 (0\%) & 21/344 (6.10\%) \\
        \gpt Dockerfile \& build output & 105/280 (37.50\%) & \textbf{4/9 (44.44\%)} & 2/16 (12.50\%) & 6/16 (37.50\%) & 13/22 (59.09\%)& 0/1 (0\%) & 130/344 (37.79\%) \\
        \midrule
        \textbf{\toolname (w/o feedback loop)} & 190/280 (67.86\%) & 2/9 (22.22\%) & 1/16 (6.25\%) & 6/16 (37.50\%) & \textbf{18/22 (81.82\%)}& 0/1 (0\%) & 217/344 (63.08\%) \\
        \textbf{\toolname (w feedback loop)} & \textbf{216/280 (77.14\%)} & \textbf{4/9 (44.44\%)} & \textbf{6/16 (37.50\%)} & \textbf{9/16 (56.25\%)} & \textbf{18/22 (81.82\%)} & 0/1 (0\%) & \textbf{253/344 (73.55\%)} \\
        \bottomrule
    \end{tabular}
    \label{table:results-docker-flakiness}
\end{table*}


\subsection{Experimental Setup} We initially identified 798 Dockerfiles exhibiting flaky behavior. Out of these, 100 Dockerfiles were reserved for \emph{repair demonstration} purposes. This left us with 698 Dockerfiles showing signs of flakiness during our nine-month longitudinal study. However, for the evaluation phase, we focused only on those Dockerfiles that continued to exhibit flakiness at the time of evaluation. Consequently, we evaluated FlakiDock on 344 Dockerfiles, as some causes of flakiness had been resolved over time.

\subsection{Baselines} 
We compare \toolname with \parfum \cite{durieuxparfum:parfum:2024}, \shipwright~\cite{shipwright:icse2021} and LLM-only prompting to repair Dockerfile flakiness. We chose \parfum as it is a recent work offering automated repairs for Dockerfile smells. \shipwright uses 13 repair-pattern rules to automatically repair Dockerfiles that fail to build. 
In contrast, other techniques, such as Hadolint~\cite{hadolint} and Binnacle~\cite{henkel:binnacle:icse2020} focus on detecting error patterns or smells and require manual intervention for their operation. For \gpt Dockerfile Only, we invoke the \gpt model to generate repairs based solely on the Dockerfile content. Following that, we include build output to provide more context for the LLM to generate repairs. 

We measure the effectiveness using the \textit{repair accuracy} metric, which represents the percentage of genuine repairs produced. A proposed Dockerfile is considered a genuine repair if its build is successful across $n$ builds.

\subsection{Results}
Table~\ref{table:results-docker-flakiness} presents the repair accuracy of each method across flakiness categories. \toolname (With Feedback Loop) achieves the highest success rate of 73.55\%, demonstrating the effectiveness of iterative refinement.
In comparison, \parfum has the lowest accuracy at 0.58\% and \shipwright achieves 5.52\%, both highlighting their limitations in addressing complex errors.
\gpt Dockerfile Only achieves 6.10\%, while GPT-4 Dockerfile \& Build Output improves significantly to 37.79\%, emphasizing the importance of incorporating build output context.



\header{Dependency-Related Errors (DEP)}
The repair accuracy for DEP errors with \parfum is 0\%, as it relies on predefined rules that fail to address the complex and dynamic nature of dependencies, such as version mismatches or missing libraries. \shipwright achieves 6.79\%, which is notable as this is the only category it can handle, highlighting its limitations in addressing broader flakiness issues. \gpt Dockerfile Only achieves 4.64\% accuracy, while \gpt Dockerfile \& Build Output improves to 37.50\% by leveraging build output context, enabling the LLM to better identify and resolve dependency issues. \toolname, without a feedback loop, achieves 67.86\% accuracy, significantly outperforming other methods. With the inclusion of feedback loop, \toolname yields the highest accuracy of 77.14\%, demonstrating the effectiveness of iterative refinement in resolving dependency-related errors.

\header{Server Connectivity Errors (CON)} \parfum achieves an accuracy of 11.11\% for addressing server connectivity errors. \gpt Dockerfile Only improves this to 22.22\%, while the inclusion of build output further increases the accuracy to 44.44\%, leveraging additional context for a more effective resolution. \toolname, incorporating its feedback loop, achieves an equivalent accuracy of 44.44\%, demonstrating performance comparable to the most effective LLM-based approach for this category.


\header{Security \& Authentication Errors (SEC)} Security and authentication errors pose significant challenges due to their complexity, often requiring nuanced approaches beyond static rules. \gpt Dockerfile achieves a 6.25\% success rate, reflecting limited effectiveness due to its reliance on general LLM capabilities without detailed context. Incorporating build output increases accuracy to 12.50\%, highlighting the value of additional contextual information. Without a feedback loop, \toolname also achieves 6.25\%, underscoring the importance of iterative refinement. With the feedback loop, \toolname attains the highest repair accuracy of 37.50\%, demonstrating its ability to effectively address these intricate issues.


\header{Package Manager-Related Errors (PMG)} \parfum achieves a modest success rate of 6.25\%, reflecting its limited ability to handle this category. \gpt Dockerfile Only performs better, with a 25\% success rate. This improvement is due to the LLM's general knowledge of package management, which enables it to address some common package-related issues based on Dockerfile content. Incorporating build output raises the accuracy to 37.50\%, as the additional context enables better understanding of package manager-related problems. \toolname, without a feedback loop, matches this 37.50\% accuracy. However, when the feedback loop is incorporated, the highest accuracy achieved is 56.25


\header{Environment Errors (ENV)} \gpt Dockerfile achieves a 4.55\% success rate, demonstrating limited effectiveness due to its reliance solely on Dockerfile content. Incorporating build output raises the success rate to 59.09\%, as the additional context allows the LLM to better understand and resolve environmental issues. \toolname outperforms all other methods, achieving an 81.82\% success rate even without a feedback loop, demonstrating its strong performance in addressing environmental errors. However, incorporating a feedback loop does not provide additional improvement.


\header{Filesystem-Related Errors (FS)} For this flakiness category, all techniques, including \toolname, fail to resolve any flakiness. The primary challenge lies in the inability of existing tools and LLMs to account for changes in the updated image's filesystem, such as a modified \code{WORKDIR}. Although feedback or similar examples could potentially aid in addressing these issues, our evaluation is limited to a single example, making it difficult to draw generalized conclusions about this category.


%% file: discussion.tex
\section{Discussion}
\header{(Non-)Deterministic Dockerfile Flakiness}
Dockerfile flakiness can manifest as deterministic or non-deterministic temporal failures, each requiring tailored repair approaches. Deterministic flakiness results in persistent and reproducible build errors caused by specific external changes, such as the deprecation of a dependency or removal of a required resource. For example, in Listing~\ref{lst:dis-deterministic}, a deterministic failure occurs due to the unavailability of the \code{npm.taobao.org} host, causing the \code{wget} command to fail. \toolname analyzes the static and dynamic build information, identifies the issue, and proposes a repair by substituting the unavailable URL with a valid one, as shown in the repaired Dockerfile. However, even after resolving a deterministic error, it is important to note that new issues may arise in the future due to the evolving nature of Dockerfile dependencies and environments. 


\begin{lstlisting}[language=Dockerfile, captionpos=t, title={\small\textbf{Build Output Summary}}, numbers=none]
wget: unable to resolve host address 'npm.taobao.org'
\end{lstlisting}
\vspace{-0.9\baselineskip}
\begin{lstlisting}[language=Dockerfile, captionpos=t, backgroundcolor=\color{highlight}, title={\small\textbf{Repaired Dockerfile}}, numbers=none]
- RUN wget https://npm.taobao.org/mirrors/node/v8.9.1/node-v8.9.1-linux-x64.tar.gz && \
+ RUN wget https://nodejs.org/dist/v8.9.1/node-v8.9.1-linux-x64.tar.gz && \
\end{lstlisting}
\vspace{-0.9\baselineskip}
\begin{lstlisting}[language=Dockerfile, label={lst:dis-deterministic}, captionpos=b, caption={Host Resolution Error (Deterministic)}, numbers=none]
  tar -C /usr/local --strip-components 1 -xzf node-v8.9.1-linux-x64.tar.gz && \
  rm node-v8.9.1-linux-x64.tar.gz
\end{lstlisting}

In contrast, non-deterministic flakiness arises sporadically due to transient factors, such as server connectivity interruptions or temporary package manager caching inconsistencies. These failures lack a consistent pattern, complicating root-cause analysis and repair validation. For instance, Listing~\ref{lst:dis1} demonstrates a timeout error caused by a transient network issue during a \code{yarn} install command. To address this, \toolname adjusts the network timeout configuration in the Dockerfile to ensure stability, as shown in the repaired version. While \toolname can propose an initial repair based on observed data, a single successful repair validation is insufficient. Instead, repairs for non-deterministic failures must be tested iteratively over time to ensure reliability and prevent reoccurrence. The repair validator in \toolname facilitates this by running repeated builds, monitoring outcomes across multiple iterations, and confirming the robustness of the repair.

\begin{lstlisting}[language=Dockerfile, captionpos=t, title={\small\textbf{Build Output Summary}}, numbers=none]
An unexpected error occurred: "https://registry.yarnpkg.com/@material-ui/icons/-/icons-4.11.3.tgz: ESOCKETTIMEDOUT".
\end{lstlisting}
\vspace{-0.9\baselineskip}
\begin{lstlisting}[language=Dockerfile, captionpos=t, title={\small\textbf{Repaired Dockerfile}}, numbers=none]
FROM node:16.14.0
...
COPY --from=deps /tmp/deps.json ./package.json
COPY yarn.lock ./
\end{lstlisting}
\vspace{-0.9\baselineskip}
\begin{lstlisting}[language=Dockerfile, captionpos=t, backgroundcolor=\color{highlight}, numbers=none]
+ RUN yarn config set network-timeout 600000
\end{lstlisting}
\vspace{-0.9\baselineskip}
\begin{lstlisting}[language=Dockerfile, label={lst:dis1}, captionpos=b, caption={Timeout Error (Non-Deterministic)}, numbers=none]
RUN yarn install --production
\end{lstlisting}

In \toolname, deterministic flakiness is repaired with greater certainty due to its consistent root cause. Non-deterministic flakiness, however, requires iterative testing and extended validation to ensure repair effectiveness. \toolname addresses both by offering immediate repairs for deterministic flakiness and incorporating a time-based validation process for non-deterministic cases, adjusting parameters such as $n$ for iterative validation.

\header{Threshold $n$ for Determining a Successful Repair}
The parameter $n$, representing the number of build iterations used to validate repairs, is crucial for ensuring repair accuracy in addressing Dockerfile flakiness. In our study, $n$ was set to two based on empirical analysis of the demonstration dataset, capturing over 90\% of flaky behaviors. For temporal deterministic errors, this value is sufficient, as these errors are resolved once the root cause is addressed. Once a deterministic error has been repaired successfully in a single build, it is unlikely to reoccur, so additional iterations have little effect on repair accuracy. 

Non-deterministic flakiness presents a more complex challenge. Since these errors can occur unpredictably, passing two rounds of validation may not suffice to reliably capture and resolve the flakiness, requiring a higher $n$ to ensure robustness. A potential enhancement to \toolname would be the inclusion of a detector that identifies the type of Dockerfile flakiness (deterministic or non-deterministic). Based on this classification, and based on the retrieved similar records, the parameter $n$ could be adjusted accordingly. For deterministic flakiness, a lower $n$ is sufficient, while non-deterministic flakiness would require a higher $n$ to ensure a reliable repair. This would prevent premature validation of repairs for non-deterministic errors.

\header{Dynamic Nature of Dockerfile Flakiness} 
Dockerfiles rely on evolving components such as base images, package managers, libraries, external URLs, and infrastructure. Changes in these dependencies such as deprecated libraries, broken URLs, or updated environment variables, can introduce flakiness, causing builds to fail. Conversely, flakiness can be resolved as dependencies stabilize without Dockerfile modifications, highlighting the temporal and dynamic nature of Dockerfile flakiness. Our nine-month study revealed the temporal nature of Dockerfile flakiness, where initially stable Dockerfiles became flaky due to dependency updates, and flaky ones stabilized as external issues were resolved. Static analysis tools fail to detect such failures, as they lack awareness of dynamic dependency changes (e.g., Listings \ref{lst:mot1} and \ref{lst:mot2}). We propose a taxonomy to help developers systematically diagnose and address flakiness while preparing for future instability. This underscores the need for \toolname, which dynamically adapts to evolving dependencies by providing context-aware repairs.

\header{Limitations of Existing Tools for Dockerfile Flakiness}
Existing tools for Dockerfile repair, such as \shipwright and \parfum, exhibit significant limitations when addressing the dynamic and multifaceted nature of Dockerfile flakiness. \shipwright achieved only 19 repairs (5.52\%) in our evaluation. 
However, this represents an upper bound, as a repair is deemed successful if \shipwright generates an output Dockerfile using its predefined patterns, irrespective of whether it fully addresses the underlying flakiness.
This tool relies on 13 repair patterns, heuristically designed by its authors through a clustering approach. These patterns must be maintained and updated to remain effective as new failure types emerge. Without regular updates, the effectiveness of the tool is likely to decline, given the dynamic and evolving nature of Dockerfile dependencies, which often introduce novel and unforeseen failure scenarios.
\shipwright was capable of handling only Dependency-Related Errors (DEP), likely because such errors often align with its predefined patterns, such as missing or incorrect dependencies. 
Similarly, \parfum focuses on repairing Dockerfile smells but lacks mechanisms to address dynamic flakiness, achieving only a 0.58\% repair accuracy. 
In contrast, our proposed tool, \toolname, integrates both static and runtime contexts using retrieval-augmented generation and an iterative feedback loop, achieving a significantly higher repair accuracy of 73.55\%. These results highlight the shortcomings of static, heuristic-driven tools and emphasize the need for dynamic, context-aware approaches to effectively handle the complexities of Dockerfile flakiness. However, \toolname is not without limitations. It relies on LLMs, which may struggle with uncommon flakiness issues that fall outside their training scope. Its effectiveness depends on the diversity and quality of the demonstration dataset, limiting performance in addressing rare error categories. Additionally, the iterative repair process, while effective, can be computationally intensive and time-consuming.

%% file: threats.tex
\section{Threats To Validity}
\header{Choice of Subjects} The choice of Dockerfiles for our study can inherently introduce a bias, potentially influencing the results of our research. To address this, we employed a dataset from a prior study~\cite{shipwright:icse2021}, encompassing 8132 Dockerfiles. However, the dataset includes Dockerfiles from repositories with ten or more stars, which one might find in popular GitHub repositories, addressing the generalizability of our findings to a broader range of Dockerfiles.

\header{Duration of the Study} The duration of our study could potentially influence our findings, especially with respect to detecting Dockerfile flakiness and temporal failures. To address this, we conducted our study over a nine-month period. This allowed us to observe the Dockerfiles over a significant period, during which updates and changes are likely to occur, potentially affecting their behavior. This extended period provided us with sufficient data to identify non-deterministic behaviors and temporal inconsistencies effectively.

\header{Host Operating System} The choice of operating system for Docker can affect the results. We used a stable version of Redhat 9 to ensure consistency, but this may limit the generalizability to other systems. Different Linux distributions, or versions, might show varying flakiness due to differences in package management and system libraries. Non-Linux hosts such as Windows or MacOS could exhibit different flakiness not captured in our study. By focusing on Redhat, we minimized environmental variability, isolating Dockerfile-specific flakiness. Future research could explore Dockerfile flakiness across diverse operating systems and versions.

\header{Repair Construction Bias} As we generate the repair dataset, we might be biased in providing the repairs in a way that helps with other repairs. This bias could stem from the tendency to create repair patterns that are more easily generalizable, potentially overlooking unique or less common solutions. To alleviate this bias, we used repairs found from existing knowledge-sharing websites such as Stack Overflow, GitHub discussions, and the official Docker website. 




%% file: related-work.tex
\section{Related Work}

\header{Test Flakiness} There is a wide array of techniques that have been proposed focusing on characterizing, detecting, and repairing flaky tests \cite{lue:flakiness:fse2014, ipflakies:python-order-dependent-flaky-tests:icse22, shi:ifixflakies-order-dependent-flaky-tests:fse2019, chen:flakydoctor:icse2024, fatima:flakyfix:arxiv2024}. Continuous integration research such as \cite{durieux:flakiness:icse2020} has a strong overlap with test flakiness literature, evaluating the prevalence and impacts of test flakiness in systems that involve test executions. Flaky tests are also a concern in user interface testing~\cite{romano:ui-based-falky-tests:icse21}. We refer to a recent survey for a more comprehensive discussion on flaky tests~\cite{parry:flaky-test-survey:tosem21}. In contrast, we are the first to examine flakiness from the perspective of Dockerfiles.


\header{Dockerfile Analysis} In Sections \ref{sec:introduction} and \ref{sec:motivation}, we discussed recent work on Dockerfile analysis \cite{giovanni:not-all-dockerfile-smells-are-same:msr24, henkel:binnacle:icse2020, zhou:drive:arXiv2022, durieuxparfum:parfum:2024, rosa:fixing:arXiv2022, shipwright:icse2021, dockercleaner:icsm2023, eng:msr2021, hadolint, dockerfilelint}. \toolname differentiates itself from existing Dockerfile smell detection tools \cite{henkel:binnacle:icse2020, hadolint, dockerfilelint, dockercleaner:icsm2023} and repair tools such as \cite{durieuxparfum:parfum:2024} by not only addressing static issues within Dockerfiles but also targeting the dynamic errors caused by the flakiness of Dockerfiles. 
\shipwright~\cite{shipwright:icse2021} introduces a human-in-the-loop approach for repairing broken Dockerfiles. \shipwright begins by clustering broken Dockerfiles to identify common failure patterns. This is achieved by embedding build log text into vector representations using a modified BERT language model, followed by clustering with the HDBSCAN \cite{HDBSCAN} algorithm. Representative Dockerfiles from each cluster are then presented to a human supervisor, who searches for solutions to the identified failures. Discovered solutions are evaluated for applicability across the cluster, and successful fixes are formalized into a database of repair patterns. Each pattern comprises a regular expression to identify issues and a corresponding repair function detailing the fix.
In contrast, \toolname is fully automated, leveraging LLMs, retrieval-augmented techniques, and a feedback loop to refine repairs without requiring manual oversight.

\header{Learning-based Program Repair} Learning-based program repair has been extensively studied in the literature~\cite{sequencer, katana, glance, reptory, yasunaga:drrepair:icml20, tang:graph2seq-for-repair:2021, ding:patching-as-translation:ase22}. Unlike these approaches, which involve training task-specific models, \toolname uses a general-purpose LLM without the need for model training.


\header{LLM-based Program Repair} There has been increasing focus on applying LLMs to program repair tasks. Early studies focused on using prompts and error messages to generate source code repair in a single interaction with the model \cite{xia:automated-repair-with-llm:icse23,  cedar:icse2023}. More recent techniques involve iterative approaches, querying the LLM multiple times and refining repairs based on feedback from previous attempts to repair source code\cite{xia:chatgpt-feedback:arxiv23}. In contrast, in this work, we leverage both static and dynamic information from Dockerfiles to provide the LLM, enhancing its ability to repair flakiness more effectively. By integrating retrieval-augmented generation techniques, we further ensure that the LLM is equipped with relevant examples and contextual knowledge, leading to more accurate and reliable repairs.






%% file: conclusion.tex
\section{Conclusion}
We present the first comprehensive study on Dockerfile flakiness, identifying that 9.81\% of Dockerfiles exhibit flaky behavior, significantly impacting the reliability of CI/CD pipelines. To address this issue, we introduce a novel taxonomy of Dockerfile flakiness and propose \toolname that leverages language models, retrieval-augmented generation, dynamic analysis, and an iterative feedback loop for automatic repair. 
Our evaluation shows that FlakiDock achieves a 73.55\% repair accuracy, outperforming \parfum, \shipwright, and GPT-4-based prompting by 12,581\%, 1,232\%, and 95\%, respectively. These results highlight the effectiveness of \toolname in addressing Dockerfile flakiness. 

\section{Data Availability}
We have made our dataset, model, comparison framework, and \toolname's implementation available~\cite{flakidock} for the reproducibility of results. 
